\documentclass[a4paper,11pt]{article}
\pdfoutput=1 
\usepackage[utf8]{inputenc}
\usepackage{float}
\usepackage{jinstpub} 
\usepackage{caption}
\usepackage{subcaption}
\DeclareUnicodeCharacter{2212}{\ensuremath{-}}
\usepackage{lineno}
\title{\boldmath Novel resistive charge-multipliers for dual-phase LAr-TPCs: towards stable operation at higher gains}

\author[a]{A. Tesi\note{Corresponding author}}
\author[a]{, L. Moleri}
\author[b]{, S. Leardini}
\author[a]{, A. Breskin}
\author[b]{, D. Gonzalez-Diaz}
\author[b]{, L. Olano-Vegas}
\author[a]{, A. Jash}
\author[a]{and S. Bressler}

\affiliation[a]{Department of Particle Physics and Astrophysics, Weizmann Institute of Science, Israel}
\affiliation[b]{Instituto Galego de Física de Altas Enerxías (IGFAE), Universidade de Santiago de Compostela, Spain}

\emailAdd{andrea.tesi@weizmann.ac.il}

\abstract{
\noindent
Cryogenic versions of Resistive  WELL (RWELL) and Resistive Plate WELL (RPWELL) detectors have been developed, aimed at stable avalanche multiplication of ionization electrons in the vapor phase of LAr (dual-phase TPC). In the RWELL, a thin resistive DLC layer deposited on top of an insulator is inserted in between the electron multiplier (THGEM) and the readout anode; in the RPWELL, a resistive ferrite plate is directly coupled to the THGEM. Radiation-induced ionization electrons in the liquid are extracted into the gaseous phase. They drift into the THGEM's holes where they undergo charge multiplication. Embedding resistive materials into the multiplier proved to enhance operation stability due to the mitigation of electrical discharges - thus allowing operation at higher charge gain compared to standard THGEM (a.k.a. LEM) multipliers. We present the detector concepts and report on the main preliminary results.

}


\begin{document}
\maketitle
\flushbottom

\section{Introduction}
\label{sec:intro}
In the last years, noble-liquid detectors have predominantly gained the scene in the field of particle and astroparticle physics as a leading technology: examples can be found in the context of neutrino physics \cite{EREDITATO, Acciarri}, dark matter searches \cite{Baudis_2012, Aprile_2010, McKinsey_2016, Aalbers_2016, Aalseth_2018, Aprile_2012, Aprile_2017} and rare events searches such as $\mu \rightarrow e\gamma$ \cite{Mihara_2011}.
A widespread concept is the dual-phase TPC \cite{Chepel_2013, Aimard_2018}: charges deposited in a liquefied noble element are extracted into its vapor phase; they induce electroluminescence (EL) photons that are detected by photo-sensors. Early attempts to multiply charges in Ar vapor through LEM amplification (targeting neutrino experiments) suffered from low charge gain (G$\sim$20 in a prototype of $\mathrm{10\times10~cm^2}$ \cite{Cantini_2015} and G$\sim$1.9 in a $\mathrm{50\times50~cm^2}$ one \cite{WA105:2021zin}, both irradiated with cosmic rays at 87~K, 0.98~bar) - due to electrical instabilities.
The possibility of enhancing the ionization signal through charge multiplication would allow a direct increase of signal-to-noise ($S/N$) compared to charge-collection in liquid phase, thus reducting the detection energy threshold.
In light of this, we propose to embed discharge-mitigating resistive materials into the detector assemblies based on Thick Gaseous Electron Multiplier (THGEM)-like concepts  \cite{Breskin_2009, BELLAZZINI, BRESKIN_2013, BRESSLER_review} in order to enhance gain and electrical stability. Following the considerable experience gained from operation at room temperature, we opted for the Resistive WELL (RWELL) \cite{Arazi_2014} and the Resistive Plate WELL (RPWELL) \cite{Rubin_2013}; they resort, respectively, to a resistive film deposited on an insulator and to a resistive plate, directly coupled to a THGEM electrode in both cases (Fig.\ref{fig:schemes}-left). A successful operation of an RPWELL detector at liquid xenon temperature (163~K) was already reported in a gaseous admixture of Ne/CH$_4$ \cite{Roy_2019}. \\

\noindent
In this work, we describe the principles of both multipliers and present, for the first time, the results of their operation in pure argon vapor at 90 K.

\section{{Experimental Setup and Methodology}}

\begin{figure}[H]
 \centering
    \includegraphics[width=12cm]{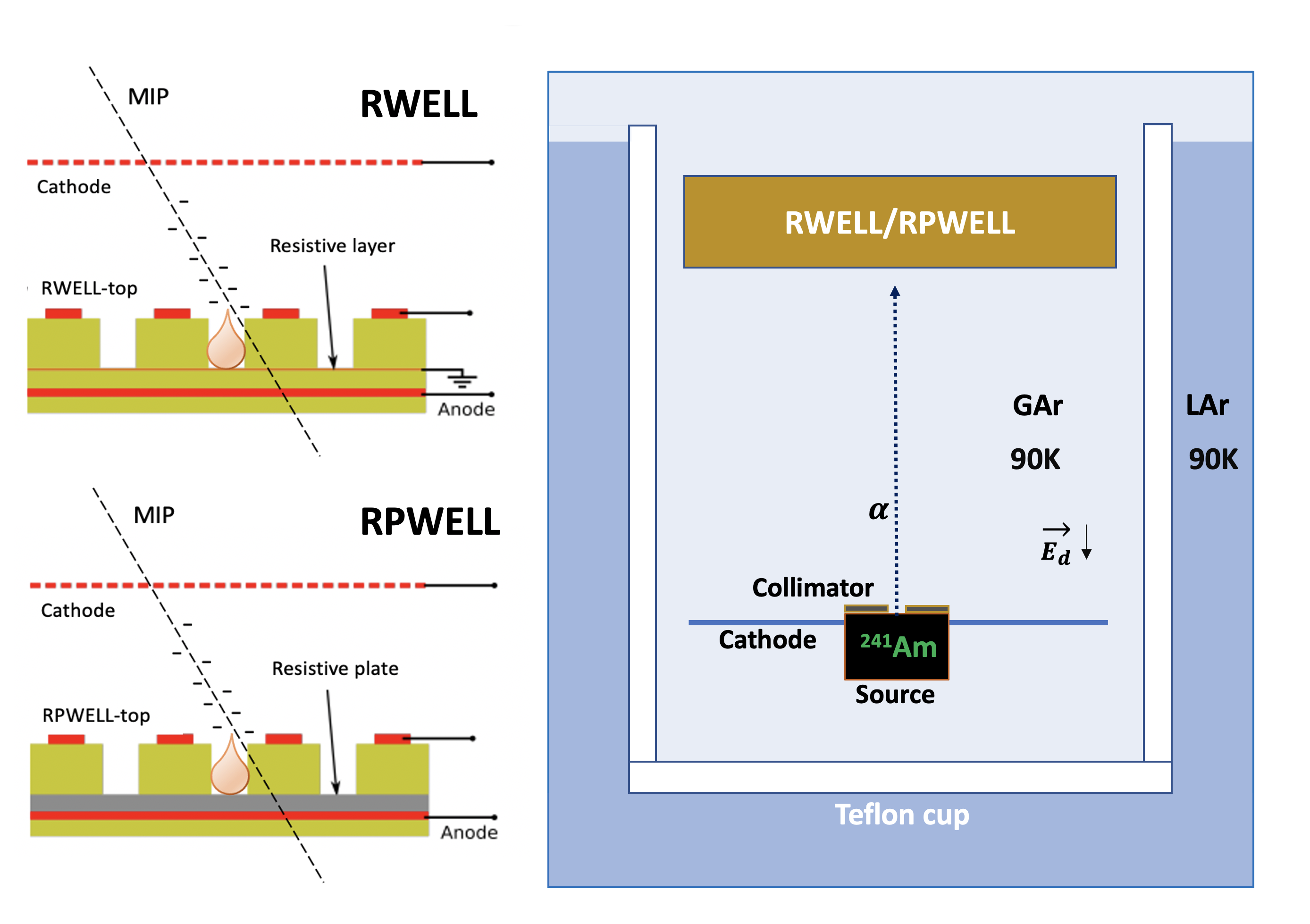}
    \caption{\textbf{(Left):} schemes of the RWELL and RPWELL detectors; \textbf{(Right):} the cryogenic setup for the operation of the RWELL/RPWELL detector in cold gaseous argon.}
    \label{fig:schemes}
\end{figure}

\noindent
In the cryo-RWELL, a perforated electrode (a 0.8~mm-thick THGEM in this case, 0.5~mm diameter holes distributed in a hexagonal pattern with 1~mm pitch, $\mathrm{3\times3~cm^2}$ in size) is screwed to the readout anode via a resistive layer of Diamond-Like-Carbon (DLC) \cite{ROBERTSON_1992} deposited on top of a 0.1~mm-thick insulator. Details about the properties of these materials and their surface resistivity at cryogenic temperature can be found in \cite{Leardini_2022}.
In the cryo-RPWELL, the multiplier is directly coupled to the anode through a resistive ceramic plate made of Yttria-Stabilized-Zirconia (YSZ) and hematite (Fe$_2$O$_3$) \cite{Morales_2013} and fixed with screws. Ceramics samples with the right resistivities (optimal bulk resistivity range: 10$^{9}$-$10^{12}~\Omega\cdot$cm) were engineered at the Instituto de Cerámica de Galicia in Spain (ICG) and characterized for operation at 90~K.
The detector assembly was inserted inside a Teflon cup immersed in the liquid argon cryostat (Fig.\ref{fig:schemes}-right), and operated in gaseous argon at 90~K. Prior to any measurement, the system was vacuum-pumped for at least 24~h reaching a pressure of $\sim$ -1$\times$10$^{-4}$~mbar. During operation, the gas was recirculated and purified with a hot getter\footnote{Entegris HotGetter PS3-MT3-R-2} in order to grant a nominal impurity content of the order of 1~ppb. The detector was irradiated with $\alpha$-particles of 5.5~MeV from an $^{241}$Am source collimated and attenuated to 4~MeV by a 12~$\mathrm{\mu m}$ mylar foil. The rate of the source after collimation was approximately 10~Hz.
Radiation-induced ionization electrons generated in the gas were transferred into the electrode’s holes. The presence of a high local electric field led to avalanche multiplication in the holes. The avalanche-charge evacuation was realized differently in the two detectors: in the RWELL, the amplified charge propagated to the ground sideways across the resistive layer; in the RPWELL, the amplified charge traveled through the resistive plate directly to the anode. 
For both processes, signals were induced on the anode and recorded with a standard electronic chain comprising a charge-sensitive preamplifier\footnote{Cremat: Model CR-110 with CR-150-R5 evaluation board}, a linear amplifier\footnote{Ortec Model 450} and an MCA\footnote{Amptek MCA 8000D}. High voltages were supplied to all the electrodes through a power supply\footnote{CAEN N1471H}. \\

\noindent
Measurements were performed by recording spectra from the charge-sensitive preamplifier, linear amplifier, and MCA electronic chain. For each voltage configuration, about 1k spectra of 120~s were acquired. Each spectrum was fitted with a Gaussian function in order to extract the mean and standard deviation of the main peak in the distribution. To estimate the gain of the detector, a measurement in collection mode was performed. The gain of the detector G was estimated by dividing the mean of the Gaussian peak in the spectrum with amplification P$\mathrm{_{Amplif}}$ by the mean value of the collection peak P$\mathrm{_{Coll}}$, $\mathrm{  G = {P_{Amplif}}/{P_{Coll}}}$. Prior to each measurement, the detector was stabilized for several hours in order to remove the charging-up effect. 
\\

\section{Results}
\label{sec:summary}
In the following, we report on some preliminary results.

\subsection{Cryo-RWELL}
\label{subsec:RWELL}

An extensive study of the properties of DLC films was conducted at cryogenic temperature and summarized in \cite{Leardini_2022}. In the current RWELL detector, a layer with surface resistivity R$_S$ equal to 20~G$\Omega/\square$ at 90~K was employed. The latter was selected after several failed attempts to stably operate the RWELL with a layer of R$_S$=10 M$\Omega/\square$, as per experience at room temperature \cite{Arazi_2014}.



\begin{figure}[H]
     \centering
     \begin{subfigure}[b]{0.497\textwidth}
         \centering
         \includegraphics[width=\textwidth]{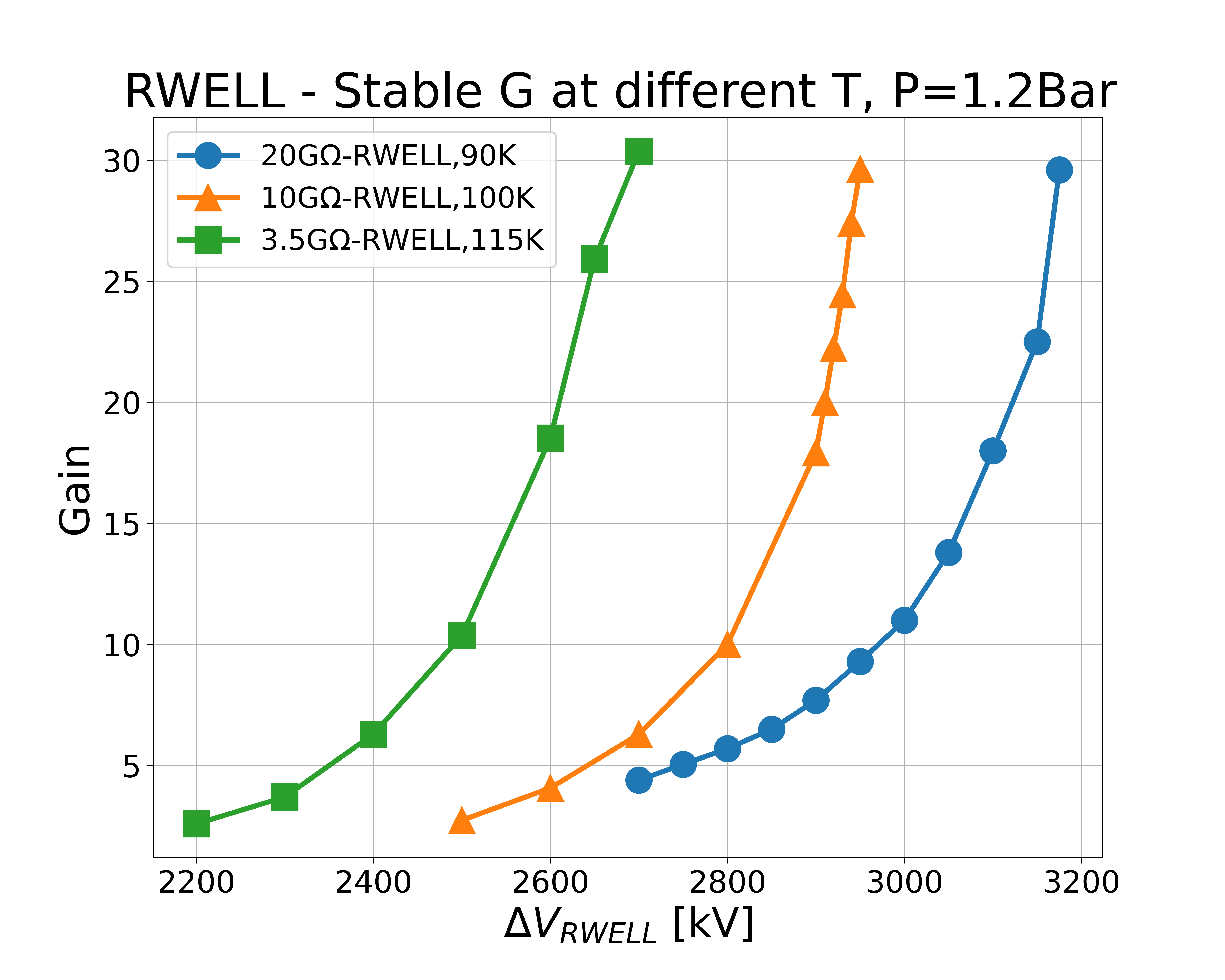}
         \caption{}
         \label{fig:RWELL_gain}
     \end{subfigure}
     \hfill
     \begin{subfigure}[b]{0.497\textwidth}
         \centering
         \includegraphics[width=\textwidth]{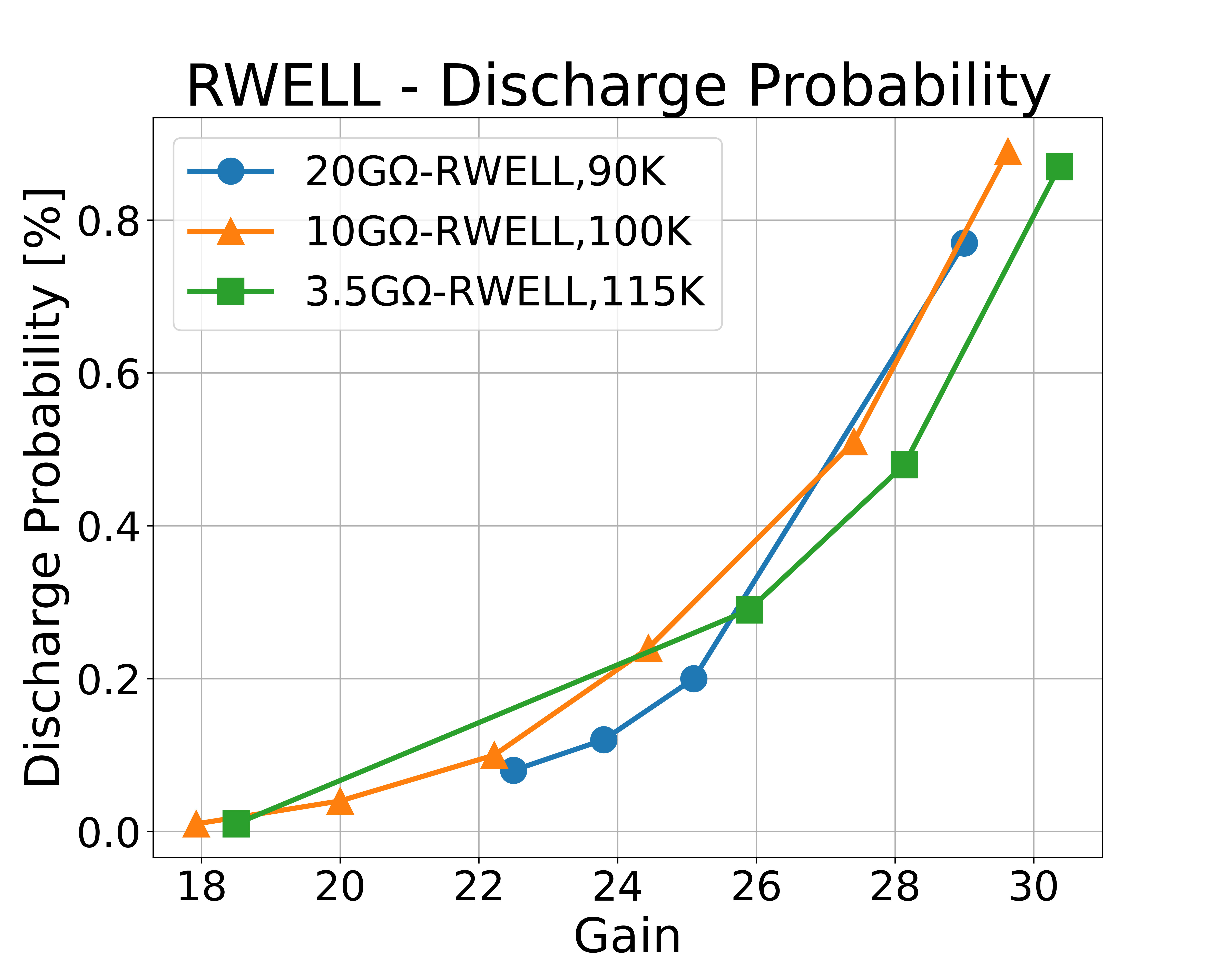}
         \caption{}
         \label{fig:RWELL_pd}
     \end{subfigure}
     \hfill
   \caption{\footnotesize \textbf{(a)}: RWELL gain as a function of the applied voltage at T=90~K, 100~K and 115~K. The resistivity of the DLC layer at each temperature is also indicated in the figure; \textbf{(b)}: discharge probability as a function of the RWELL gain for the three temperature values. } 
    \label{fig:all_gain}
\end{figure}

\paragraph{Gain $\&$ Discharge Probability}
The RWELL gain characterization was performed at three different temperatures: 90~K, 100~K and 115~K. The warming-up was performed in a controlled way to ensure thermal stability. The pressure inside the cryostat was kept at 1.2~bar throughout the entire study. The drift field was kept constant, E$\mathrm{_d}$= 0.5 kV/cm. It is important to mention that the DLC resistivity decreased while increasing temperature, from 20 G$\Omega/\square$ at 90 K to 10 G$\Omega/\square$ at 100 K till 3.5 G$\Omega/\square$ at 115 K.
The RWELL gain as a function of $\Delta$V$\mathrm{_{RWELL}}$ at different temperatures is reported in Fig.\ref{fig:RWELL_gain}. 

\noindent
The three curves manifest the same exponential trend, suggesting the presence of avalanche multiplication. A maximum stable gain close to 30 was achieved in the three cases. The left-shift at higher temperature is due to the lower gas density. Above a detector gain close to 18, the presence of occasional discharges was observed. The detector was able to sustain them and its performance was not degraded. Above 3.2~kV, the presence of self-sustained currents across the amplification stage prevented further operation.
The discharge probability is defined here as the number of discharges measured in a specific time frame, relative to the the rate of alpha particles passing through the collimator. In Fig.\ref{fig:RWELL_pd}, the discharge probability as a function of gain is depicted, for different temperatures and resistivity values. It is worth noting that the probability trend is very similar despite the different temperature conditions and a $\sim$1 $\%$ discharge probability is achieved at the maximum voltage for all three cases. This is in line with room temperature results \cite{Jash_2022}.\\ 

\subsection{Cryo-RPWELL}
\label{subsec:RPWELL}
\begin{figure}[H]
    \centering
    \includegraphics[width=9cm]{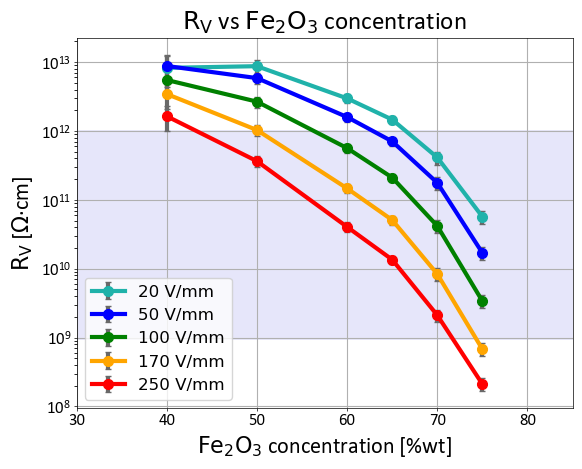}
    \centering
    \caption{Characterization of the bulk resistivity of the YSZ/Fe$_2$O$_3$ ceramics as a function of the concentration of Fe$_2$O$_3$ by weight, at LN$_2$ temperature (77~K). The shaded band represents the experimentally-motivated region where resistive materials are known to quench the spark development.
    }
    \label{fig:Rv_conc}
 \end{figure}

The availability of novel ceramic materials with a bulk resistivity R$_V$ comprised in the range 10$^9$-10$^{12}~\Omega\cdot$cm at 90~K paved the way for the construction of the cryo-RPWELL. In fact, values of R$_V$ in this range were found to be sufficient to fully quench discharges at room temperature \cite{Jash_2023}. In Fig.\ref{fig:Rv_conc}, the R$_V$ of several ceramic plates is reported as a function of  Fe$_2$O$_3$ concentration for different applied fields at 77~K. The resistive plate in the RPWELL detector over the relevant $\Delta V_{RPWELL}$ range is expected to be at ground and get minimally affected by the presence of discharges. \\

 \noindent

\begin{figure}[H]
     \centering
     \begin{subfigure}[b]{0.495\textwidth}
         \centering
        \includegraphics[width=\textwidth]{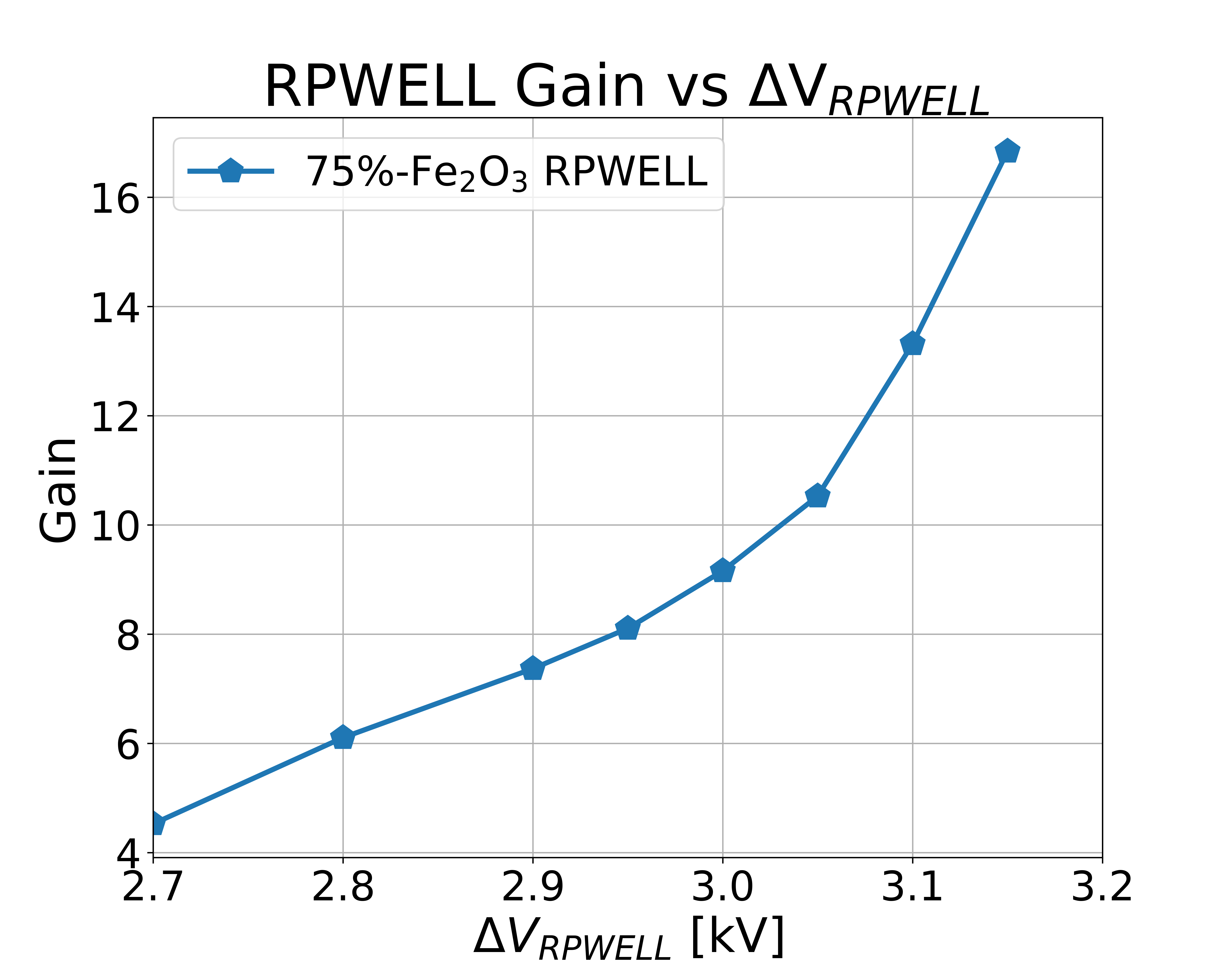}
         \caption{}
         \label{fig:RPWELL_gain}
     \end{subfigure}
     \hfill
     \begin{subfigure}[b]{0.493\textwidth}
         \centering
         \includegraphics[height=0.8\textwidth, width=1.1\textwidth]{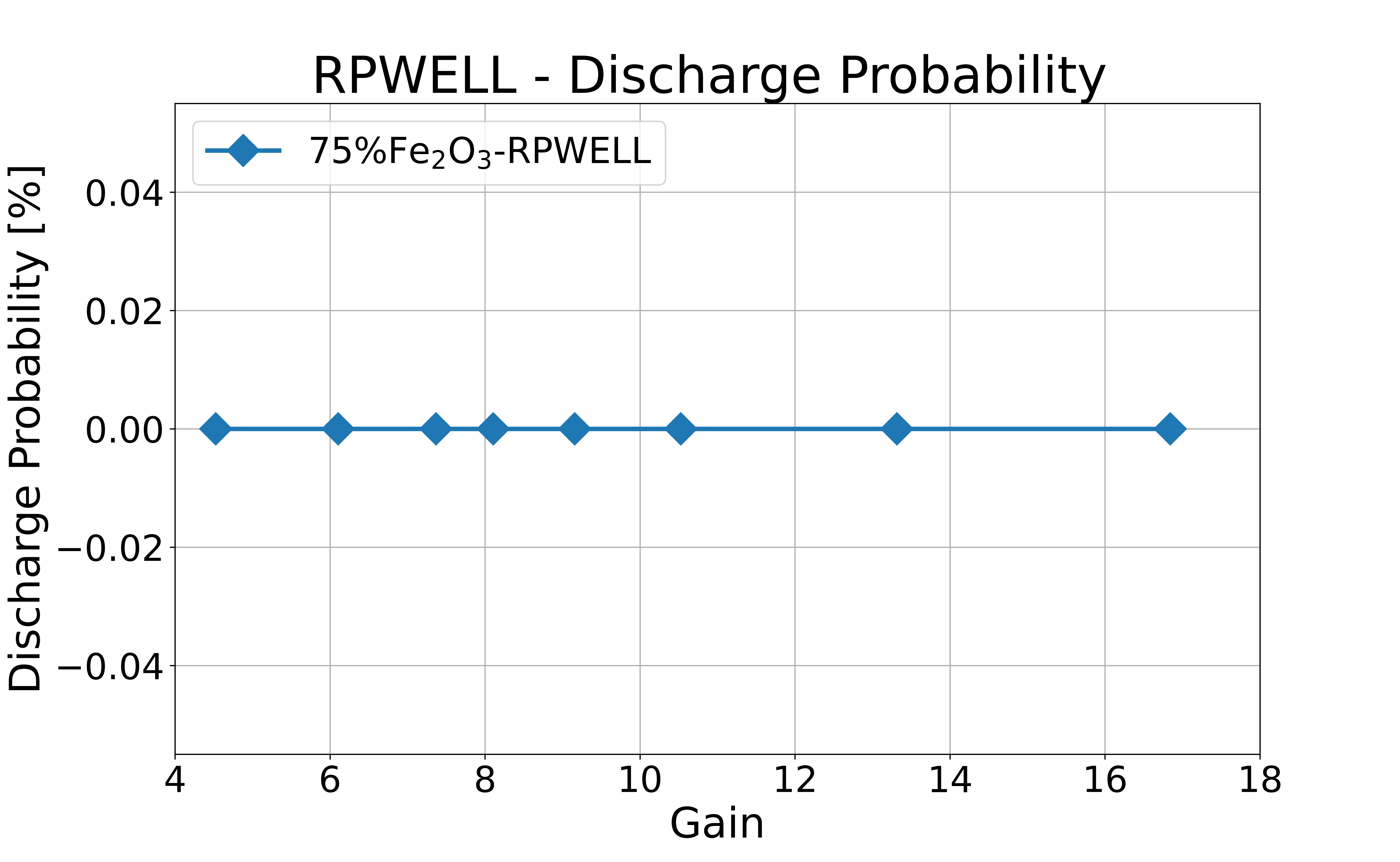}
         \caption{}
         \label{fig:RPWELL_pd}
     \end{subfigure}
     \hfill
   \caption{\footnotesize \textbf{(a)}: a stable gain curve for alpha particles recorded with the RPWELL  for 2.7 kV <$\Delta$V$_{RPWELL}$ < 3.15~kV at 90~K, 1.2~bar.; \textbf{(b)}:  discharge probability as a function of the RPWELL gain at 90 K, 1.2 Bar.   } 
    \label{fig:all_gain}
\end{figure}

\noindent
\noindent
The voltage applied across the multiplier was scanned in the range of $\Delta$V$_{RPWELL}$=2.7-3.15~kV. 
The stable gain curve with alpha particles at 90~K, 1.2~bar is given in Fig.\ref{fig:RPWELL_gain}. The discharge probability as a function of gain at 90~K is reported in Fig.\ref{fig:RPWELL_pd}. 

\noindent
It is possible to see that the RPWELL is able to operate with a stable gain close to 16 in discharge-free mode.
The investigation of this concept is still ongoing.\\

\section{Discussion}
\label{sec:discussion}

The cryo-RWELL and cryo-RPWELL concepts, incorporating resistive anodes, may evolve in the future into viable charge sensors for large-volume dual-phase radiation detectors. These concepts have been validated with small prototypes in the vapor phase of LAr. \\

\begin{figure}[H]
    \centering
    \includegraphics[width=9cm]{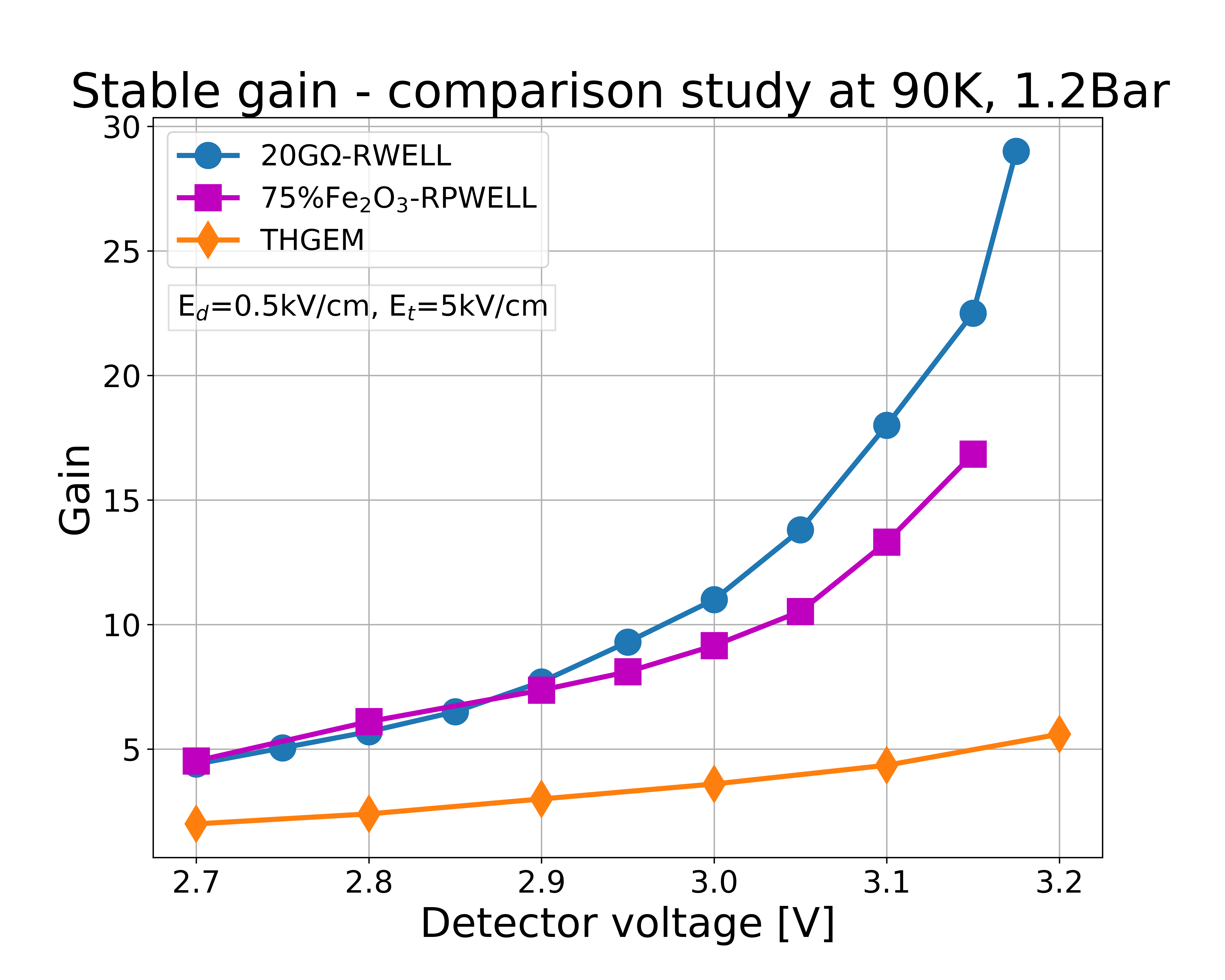}
    \centering
    \caption{Stable gain curves for different charge-amplification concepts at 90~K and 1.2~bar: THGEM+2mm induction gap (orange), 20G$\Omega/\square$-RWELL (blue) and 75$\%$Fe$_2$O$_3$-RPWELL (purple). In all cases, the THGEM stage was 0.8~mm-thick. }
    \label{fig:Comparison}
 \end{figure}

\noindent
A comparison with a standard 0.8 mm thick THGEM followed by 2 mm induction gap was also carried out, see Fig.\ref{fig:Comparison}.

\noindent
These preliminary results show that embedding resistive anodes into the detector improves its performance and protects the electronics from electrical instabilities. A detailed study of the maximal stable gain achieved by the RWELL/RPWELL is currently ongoing. As an observation, the gain of resistive anodes is 2.5-5 times higher than the gain of a THGEM operated in the same setup under the same conditions. Although a similar gain was observed in 10x10cm$^2$ LEM operated in LAr vapor with no resistive materials, a direct comparison cannot be made due to possible differences in operating conditions and gas purity; considerations about the scalability of DLC layers vs resistive plates will be addressed in a future study.  
In conclusion, while showing promising results for both RWELL/RPWELL, the data presented here raises interesting questions regarding the role played by the resistive materials in the discharge quenching process - e.g., is there a difference in the quenching mechanism between the RWELL and the RPWELL? Further studies will follow.


\section{Acknowledgements}
\label{sec:ackowledgements}
This work was supported by Sir Charles Clore Prize, by Martin Kushner Schnur, by the Nella and Leon Benoziyo Center for High Energy Physics, and by RD51 funds through its ‘common project’ initiative. We acknowledge as well financial support from Xunta de Galicia (Centro singular de investigación de Galicia, accreditation 2019- 2022), and by the “María de Maeztu” Units of Excellence program MDM-2016-0692. DGD was supported by the Ramón y Cajal program (Spain) under contract number RYC-2015-18820.

\bibliographystyle{JHEP}
\bibliography{bibliography}

\end{document}